\documentclass[12pt]{article}
\usepackage{amssymb}
\usepackage{amsmath,,calrsfs}
\usepackage[dvips]{epsfig}

\newcommand{\be}{\begin{equation}}
\newcommand{\ee}{\end{equation}}
\def\bea{\begin{eqnarray}}
\def\eea{\end{eqnarray}}

\newcommand{\bn}{\begin{eqnarray}}
\newcommand{\en}{\end{eqnarray}}

\newcommand{\p}{\partial}

\newcommand{\nn}{\nonumber}

\newcommand{\no}{\noindent}

\newcommand{\s}{\,\,\,\,}
\def\bea{\begin{eqnarray}}
\def\eea{\end{eqnarray}}

\newcommand{\beq}{\begin{eqnarray}}
\newcommand{\eeq}{\end{eqnarray}}
\begin{document}

\title{\textbf{Reaching out a ``geometrical'' description for spin-4 self-dual models in $D=2+1$ }}
\author{Elias L. Mendon\c ca \footnote{eliasleite@feg.unesp.br}, H. L. Oliveira\\
\textit{{UNESP - Campus de Guaratinguet\'a - DFI} }\\
\textit{{Av. Dr. Ariberto Pereira da Cunha, 333} }\\
\textit{{CEP 12516-410 - Guaratinguet\'a - SP - Brazil.} }\\
}
\date{\today}
\maketitle

\begin{abstract}
Starting with a first order in derivatives self-dual model which describes a massive spin-4 mode in $D=2+1$ dimensions, we have obtained a sequence of three more new descriptions, which then give us an interconnected self-dual chain  $SD(i)$ with $i=1,2,3,4$ indicating the order in derivatives. We have demonstrated that a powerful notation in terms of a self-adjoint operator $\Omega$ in the frame-like scenario truly simplifies the investigation for new models and at the third order level can be converted to a geometrical description in terms of the much more usual totally symmetric double traceless field. 
\end{abstract}
\newpage

\section{Introduction}

The present paper is devoted to the study of massive higher spin gauge theories in $D=2+1$ dimensions. We carry out our discussion analyzing the very first genuine example of  a higher spin field, the spin-4. Such choice is based mainly on two motivations: First, the study of planar gauge field theories and their equivalences is partially well understood for lower spins, and can give us several hints on further higher spin steps. Second, once the spin-4 field is equipped with all the higher spin Fierz-Pauli constraints (i.e: Field totally symmetric, double traceless and transverse) one could figure out if the results can be generalized for systems truly arbitrary with spin $s$. 

The reason for working with the specific planar world is related to the very interesting existence of the so called self-dual models in such dimension. This kind of model describe parity singlets of spins $+s$ or $-s$. The important thing about such models is that they are not unique, in the sense that there exists different descriptions, interconnected and equivalents via dualization procedures. Under a very interesting perspective, the self-dual models can be viewd as building blocks to the construction of partity preserving massive gauge theories in $D$ dimensions, this can be done through the so called soldering procedure, see \cite{stone, wotzasek}, through our previous experience with the lower spin cases one would obtain the Fronsdal action by soldering lower derivatives self-dual models \footnote{Here a very interesting issue can be maybe answered: Starting with the Fronsdal action resulting from the soldering process can we apply the Noether procedure in order to fulfill the table-1 of \cite{joung} on the hieararchy of higher derivative actions of higher spin fields? We think this is possible as we have demonstrated for example in \cite{nsd} for the lower spin case.}. At the same time, but in opposite direction, the self-dual descriptions can be obtained from massless theories in $D=3+1$ via Kaluza-Klein dimensional reduction \cite{rindani, dalmazidr}.

Despite the dualities involving the self-dual descriptions perhaps the most known example is the equivalence between the Maxwell-Chern-Simons model and the Self-dual model describing massive spin-1 modes in $D=2+1$ \cite{ad1,djt}. Generalizations of these discussion for other spins $s=3/2, 2, 5/2, 3$ and $4$ are also interesting suggestions \cite{deserkay,deser3/2,ak1,desermac, Aragonedeser, aragones31}, and the research for their equivalences reveals remarkable features, specially regarding the spin-2 context \cite{nsd}, because of its obvious relation with the study of massive gravity in $D=2+1$ \cite{BHT}. Besides, another interesting topic is related to the constructive and natural emergence of ``geometrical objects'' from theories originally formulated in the frame-like. This is precisely one of the issues addressed here.

In this work we have used the first order self-dual model, suggested in \cite{aragones31}, as the starting point for obtaining new gauge invariant higher order in derivatives descriptions which can be completely expressed in terms of geometrical objects like the Einstein tensor. The models obtained are complete in the sense that they contemplate the auxiliary fields needed to eliminate spurious degrees of freedom. The tool we have used for obtaining such descriptions consists of the quite well tested procedure called the Noether Gauge Embedment $NGE$, see \cite{anacleto} for an introduction. Through this procedure we have performed three rounds of gauge impositions in order to achieve a fourth order gauge invariant self-dual model. An important byproduct of the study of these free theories under such techical procedure is the trade between the auxiliary fields in higher derivatives and gauge symmetries along the proccess. Once one can demonstrate that the higher derivatives does not implies ghosts one can use the gauge symmetries as a guiding principle for the introduction of interactions, see \cite{Zinoviev} for example, where the authors study the construction of cubic vertex for massless and massive higher spin particles in $D$ dimensions in an electromagnetic background, there, in order to provide gauge invariance they have used Stuecklberg fields.  All along the work
we have noticed that there are several similarities with the previous lower spin cases \cite{nges3} (and references therein), this, in some sense, reinforces the robustness of the dualization method \footnote{Notice however that, in \cite{nsd} we show that in some conditions, which are not witnessed here, we obtain examples when the NGE fails to produce a physical gauge theory. Indeed the appearance of ghosts via	NGE has been noticed before in \cite {baeta}.}. 

The basic spin-4 field in the first order self-dual model is a generalized dreibein field  $\omega_{\mu(\alpha\beta\gamma)}$ with four index, where those which are between parenthesis are symmetric and traceless. Here we have demonstrated that the suggestion of a second order, self-adjoint operator, we have called $\Omega_{\mu(\alpha\beta\gamma)}$ is very useful for obtaining new models. We have also demonstrated that, by getting to third stage of gauge imposition, the frame like description can be completely converted in terms of a geometrical one, where the fields are then totally symmetric and the useful second order operator is automatically substituted by the relevant second order self-adjoint Einstein tensor. 

Despite the results we have obtained here, an issue still persists: Looking back to the lower spin cases we have observed for spins $1, 3/2$ and $2$ that the number of self-dual descriptions corresponds to $2s$. Apparently when we address the transition spin-3 case and subsequently the spin-4 case, this rule seems to be broken. Besides the four descriptions we connect and provide in this paper, it has been suggested recently the highest order models in \cite{ddalrs, marija}, those descriptions are also geometrical and written in terms of totally symmetric non double traceless fields, surprisingly they do not depend on auxiliary fields and despite this are free of ghosts. So far, we can not see how and if, it would be possible the connection between the models obtained here and such highest order descriptions.

\section{From $SD(1)$ to $SD(2)$}

In this section we are going to recover the first order self-dual model for massive spin-4 particles in $D=2+1$ dimensions which was suggested in \cite{aragones31}. In such model the massive mode is described in terms of a partially symmetric tensor given by $\omega_{\mu(\beta\gamma\lambda)}$ where the set of indices between parenthesis are symmetric and traceless in such a way that $\eta^{\beta\gamma}\omega_{\mu(\beta\gamma\lambda)}=\eta^{\beta\lambda}\omega_{\mu(\beta\gamma\lambda)}=\eta^{\gamma\lambda}\omega_{\mu(\beta\gamma\lambda)}=0$. On the other hand we are going to have non null trace if, for example, $\eta^{\mu\gamma}\omega_{\mu(\beta\gamma\lambda)}=\omega_{\beta\lambda}$, in other words  if the trace is taken with one indice inside and another outside of the parenthesis. The first order self-dual spin-4 action can be written as:

\begin{eqnarray}
	S_{SD(1)}&=&\int d^3x \ \ \bigg[\frac{m}{2}\epsilon^{\rho\mu\nu}\omega_{\rho(\alpha\beta\gamma)}\partial_\mu \omega_\nu^{\ \ (\alpha\beta\gamma)}- \frac{m^2}{2}\epsilon^{\rho\mu\nu}\epsilon^{\alpha\beta}_{\ \ \ \gamma}\eta_{\rho\alpha}\omega_{\mu(\beta\varsigma\lambda)}\omega_\nu^{\ \ (\gamma\varsigma\lambda)} +  m^2\omega_{\alpha\beta}U^{\alpha\beta}\bigg] +S^1_{aux} \nn \\ \label{ad1}
\end{eqnarray}
in the action (\ref{ad1}), the spin-4 field is coupled to the auxiliary rank-2 field $U^{\alpha\beta}$ and at the end of the expression we have an auxiliary action \footnote{Notice that the scalar field $U$ can be eliminated with the help of its equations of motion, in such a way that $U=9 \p_{\mu}v^{\mu}/44m$, then:\be -\frac{9m}{5}U\p_{\mu}v^{\mu}+\frac{22m^2}{5}U^2= - \frac{81}{440}(\p_{\mu}v^{\mu})^2\ee} , which can be written explicitly as:

\begin{eqnarray}
	\label{acao21}
	S^1_{aux}&=&\int d^3x  \ \ \bigg[-\frac{3m}{4}\epsilon^{\rho\mu}_{\ \ \nu}U_{\rho\alpha}\partial_\mu U^{\nu\alpha} -\frac{3m^2}{2}\epsilon^{\rho\mu\nu}\epsilon^{\alpha\beta\gamma}\eta_{\rho\alpha}U_{\mu\beta}U_{\nu\gamma}-\frac{8m}{9}\epsilon^{\mu\nu\beta}H_\mu\partial_\nu H_\beta\cr\cr
	&-& \frac{9m}{20}\epsilon^{\mu\nu\beta}V_\mu\partial_\nu V_\beta + \frac{32m^2}{9}H_\mu H^\mu - \frac{9m^2}{5}V_\mu V^\mu + m^2H^\mu V_\mu-\frac{9m}{5}U\partial_\mu V^\mu + \frac{22m^2}{5} U^2\bigg],\nn\\
\end{eqnarray}

While the field $U^{\alpha\beta}$ guarantees that no spin-2 ghosts are propagating, through the analysis of the equations of motion of the pure spin-4 part of the action (\ref{ad1}) the authors in \cite{aragones31} conclude that we still have the propagation of spin-1 and 0 variables, that is precisely why we have to consider the auxiliary vectors $H^{\mu}$ and  $V_{\mu}$ and an additional auxiliary scalar $U$ which is indeed the trace of $U^{\alpha\beta}$  \footnote{The the rank-2 field $U_{\alpha\beta}$, has the following algebraically irreducible representation: $U_{\alpha\beta}=\tilde{U}_{(\alpha\beta)}+\eta_{\alpha\beta}U/3+ \epsilon_{\alpha\beta\gamma} H^{\gamma}$. Then the vector $H_{\mu}$ actually comes from the antisymmetric part of this field. We will also see that, when deriving the equations of motion to the spin-4 field, we end up only with the symmetric traceless part of $U_{\alpha\beta}$.}. Only the determination of each numerical coefficient  by itself in the auxiliary action deserves a whole paper. Our point here is: Once we have the first order action, with all the correct coefficients, can we obtain higher order lagrangians, gauge invariant, through a rigorous and systematic way without making a complete and complicated analysis of the equations of motion. Indeed this is possible through different dual procedures. Here, we explore such idea making use of the Noether Gauge Embedment procedure.

It is very useful to define a collaborative notation to help us in the presence of too many indices and fields. We suggest for example that $\epsilon^{\mu\nu\alpha}\p_{\alpha}\equiv E^{\mu\nu}$, while $E^{\mu\nu}\omega_{\nu}^{\,\,\,(\beta\gamma\lambda)}\equiv \xi^{\mu(\beta\gamma\lambda)}$. In order to check gauge invariance of the action we think that it is better to rewrite the mass term by opening the product of Levi-Civita symbols. With all the rearrange the action from now on, we be written as:

\be S_{SD(1)}=\int d^3x \ \ \bigg[-\frac{m}{2}\omega_{\mu(\alpha\beta\gamma)}\xi^{\mu(\alpha\beta\gamma)}+ \frac{m^2}{2}(\omega_{(\mu\beta)}\omega^{(\mu\beta)}-\omega_{\mu(\alpha\beta\gamma)}\omega^{\alpha(\mu\beta\gamma)}) +  m^2\omega_{\alpha\beta}U^{\alpha\beta}+ {\cal L}^1_{aux}\ \bigg]. \label{adr}\ee

The Chern-Simons like term in the action (\ref{adr}) (the first one) is invariant under the gauge transformation: 
\begin{equation}
	\label{sim1}
	\delta_{\xi} \omega_{\mu(\alpha\beta\gamma)}=\partial_\mu\tilde{\xi}_{(\alpha\beta\gamma)},
\end{equation}
where the gauge parameter is totally symmetric and traceless.  However, it is straightforward to check that such transformation does not correspond to a symmetry of the action due to the presence of the mass and linking terms, i.e.: $\sim m^2 \omega^2 + \omega U$. We would like to become the whole action invariant under the gauge transformation (\ref {sim1}). In order to do that we proceed by defining what we will call the Euler tensor, which can be calculated through:

\bea  K^{\mu(\beta\gamma\lambda)}&\equiv& \frac{\delta S_{SD(1)}}{\delta \omega_{\mu(\beta\gamma\lambda)}}\\
&=& -m \xi^{\mu(\beta\gamma\lambda)} + \frac{m^2}{3} (\eta^{\mu\beta}\omega^{(\gamma\lambda)}+\eta^{\mu\gamma}\omega^{(\beta\lambda)}+\eta^{\mu\lambda}\omega^{(\gamma\beta)}-\omega^{\beta(\mu\gamma\lambda)}-\omega^{\gamma(\mu\beta\lambda)}-\omega^{\lambda(\mu\gamma\beta)})\nn\\
&+& \frac{m^2}{3}f^{\mu(\beta\gamma\lambda)}(\tilde{U}),\label{K}\eea

\no where a partially symmetric-traceless combination of the auxiliary fields $\tilde{U}^{(\beta\gamma)}$ is codified in the function $f^{\mu(\beta\gamma\lambda)}(\tilde{U})$ which is explicitly given by:
\be f^{\mu(\beta\gamma\lambda)}(\tilde{U}) = \eta^{\mu\beta}\tilde{U}^{(\gamma\lambda)}+\eta^{\mu\gamma}\tilde{U}^{(\beta\lambda)}+\eta^{\mu\lambda}\tilde{U}^{(\gamma\beta)}- \frac{2}{5} (\eta^{\beta\gamma}\tilde{U}^{(\mu\lambda)}+\eta^{\beta\lambda}\tilde{U}^{(\mu\gamma)}+\eta^{\gamma\lambda}\tilde{U}^{(\mu\beta)}).\ee 

As part of the systematic procedure we define a first iterated action by considering an auxiliary field $a_{\mu(\beta\gamma\lambda)}$ \footnote{ The terminology here may be cause of confusion, but differently of the auxiliary fields of the action,  $a_{\mu(\beta\gamma\lambda)}$ have nothing to do with the elimination of spurious degrees of freedom, indeed it is added only as part of the procedure of the embedding of the equations of motion and consequently of the gauge symmetry.}, which give us:

\begin{equation}
	\label{ansatz1}
	S'= S_{AD(1)} - \int d^3x\ \ \big[a_{\rho(\alpha\beta\gamma)}K^{\rho(\alpha\beta\gamma)}\big],
\end{equation}
It is worth mentioning that the auxiliary field has the same symmetry characteristics of the original field $\omega_{\mu(\beta\gamma\lambda)}$, besides, there is a prerequisite with respect to its gauge transformation which  is set to satisfy:
\be \delta_{\xi} \omega_{\mu(\alpha\beta\gamma)}=\delta_{\xi} a_{\mu(\alpha\beta\gamma)}.\ee

\no Then, taking the gauge transformation of the first iterated action $S'$ we are going to have:

\begin{eqnarray}
\delta_{\xi} S'= \int d^3x\ \ \big[-a_{\rho(\alpha\beta\gamma)}\delta_{\xi} K^{\rho(\alpha\beta\gamma)}\big]=\int d^3x\ \ \delta_{\xi} \left[\frac{m^2}{2}\big(a^{\beta\gamma}a_{\beta\gamma}-a^{\beta(\rho\alpha\gamma)}a_{\rho(\alpha\beta\gamma)}\big)\right].
\end{eqnarray}

With these result in hand one can conclude that the action defined by:
\begin{eqnarray}
S''= S'-\frac{m^2}{2}\int d^3x\ \ \big(a^{\beta\gamma}a_{\beta\gamma}-a^{\beta(\rho\alpha\gamma)}a_{\rho(\alpha\beta\gamma)}\big),\label{S2}
\end{eqnarray}
\no is, by construction, automatically gauge invariant under the transformation (\ref{sim1}), in other words $\delta_{\xi}S''=0$. The next step is to eliminate the auxiliary field by making use of its equations of motion, which give us:

\begin{eqnarray}
&&K^{\rho(\alpha\beta\gamma)}- \frac{m^2}{3}(\eta^{\rho\alpha}a^{\beta\gamma} + \eta^{\rho\beta}a^{\gamma\alpha} + \eta^{\rho\gamma}a^{\alpha\beta}-a^{\alpha(\rho\beta\gamma)} - a^{\beta(\rho\gamma\alpha)} -a^{\gamma(\rho\alpha\beta)})=0.
\end{eqnarray}
The inversion of the field $a^{\gamma(\rho\alpha\beta)}$ in terms of $K^{\rho(\alpha\beta\gamma)}$ is quite tedious, and after some manipulations one can obtain:

\be a_{\rho(\alpha\beta\gamma)} = \frac{2}{m^2}\Omega_{\rho(\alpha\beta\gamma)}(K)\quad;\quad a_{\beta\gamma}=\frac{2}{m^2}\Omega_{\beta\gamma}(K),\label{aom}\ee

\no where we have defined the very useful symbol or notation $\Omega^{\rho(\alpha\beta\gamma)}$, which is applicable to different objects and has also similarly been defined in the lower spin cases, for example for $s=3/2$ \cite{nge32}, $s=2$ \cite{ dualdesc} and $s=3$ \cite{ nges3}. Here, for the Euler tensor it is explicitly given by:
\bea  \Omega_{\rho(\alpha\beta\gamma)}(K)&\equiv&K_{\rho(\alpha\beta \gamma)}-\frac{1}{2}\Big(K_{\alpha(\rho\beta \gamma)}+K_{\beta(\rho\alpha\gamma)}+K_{\gamma(\rho\beta\alpha)}\Big)\nn\\
&-&\frac{1}{8}\Big(\eta_{\rho\alpha}K_{\beta \gamma}+\eta_{\rho\beta}K_{\alpha\gamma}+\eta_{\rho\gamma}K_{\beta\alpha}\Big)\cr\cr
&+&\frac{1}{4}\Big(\eta_{\beta\alpha}K_{\rho\gamma}+\eta_{\gamma\beta}K_{\alpha\rho}+\eta_{\alpha\gamma}
K_{\beta\rho}\Big).\label{OM}\eea

The reader can easily check its trace obtaining $ \Omega_{\beta\gamma}(K)=3K_{\beta\gamma}/8$.  After some manipulation, the substitution of the auxiliary field $a_{\rho(\alpha\beta\gamma)}$ as given by (\ref{aom}) in (\ref{S2}) one can demonstrate that:

\begin{eqnarray}
	S''= S'-\frac{1}{m^2}\int d^3x\ \ K_{\mu(\beta\gamma\lambda)}\Omega^{\mu(\beta\gamma\lambda)}(K).\label{S''}
\end{eqnarray}

At this point we can finally perform the last complicated substitution, that of the Euler tensor given by (\ref{K}) in (\ref{S''}). Such manipulation, taking in account the useful ``self-adjoint'' property of the $\Omega$-symbol, i.e.: $A\Omega(B)=B\Omega(A)$, for arbitrary $A$ and $B$ will leave us with the so called second order self-dual model which we explicitly write below:

\begin{eqnarray}
\label{Sfim}
S_{SD(2)}&=&\int d^3x\ \ \bigg[-\xi_{\rho(\alpha\beta\gamma)}\Omega^{\rho(\alpha\beta\gamma)}(\xi)
-\frac{m}{2}\omega_{\rho(\alpha\beta\gamma)}\xi^{\rho(\alpha\beta\gamma)}\nn\\
&+&\frac{3m}{4}\xi_{\beta\gamma}\tilde{U}^{(\beta\gamma)}-\frac{11m^2}{40}\tilde{U}_{(\alpha\beta)}\tilde{U}^{(\alpha\beta)}+{\cal L}^1_{aux}\bigg].\label{s2}
\end{eqnarray}
On the result obtained in (\ref{s2}) some comments are in order. First thing we observe is the emergence of a second order in derivative term, the first one in the expression, such term consists of the sum of three pieces which thanks to the $\Omega$-symbol can be arranged in a single one. Next we give the explicit expression for it. One can also notice that the Chern-Simons like term persists in the second order model, but now with a change in its sign. We have also a new linking-term between the spin-4 field and the auxiliary field $\tilde{U}_{(\alpha\beta)}$ which now is of first order in derivative, see $\sim \xi \Omega(f)$. Lastly we mention that the auxiliary lagrangian has received through the process of embedment a new contribution, the term $\sim \tilde{U}^2$, and from now on we will incorporate it by redefining the auxiliary lagrangian by ${\cal L}^{1}_{aux}\to {\cal L}^{2}_{aux}$ . Through our previous experience dealing with a similar procedure in the much simpler case of spin-3, such automatic corrections of the auxiliary lagrangian are indeed required to the maintenance of the correct spectrum of the theory, here, we are not going to focusing in demonstrating that.   One can observe that all the terms which were breaking gauge invariance of the action (\ref{adr}) have completely gotten rid off in (\ref{s2}), as a consequence, one can check that the remaining terms are in fact gauge invariant. As far as we know the second order action obtained here is an original result.

\section{From $SD(2)$  to $SD(3)$}

It turns out that, similarly to what happens in the lower spin cases, the second order term can be written in terms of a totally symmetric field $\phi_{\mu\alpha\beta\gamma}$ as we will see in the last section. This is an indication that this term is invariant under the following gauge transformation:
\begin{eqnarray}
\label{Sim2}
\delta_{\tilde{\Lambda}} \omega_{\nu(\alpha\beta\gamma)}=\epsilon_{\nu\alpha}^{\ \ \ \ \rho}\tilde{\Lambda}_{(\rho\beta\gamma)}+\epsilon_{\nu\beta}^{\ \ \ \ \rho}\tilde{\Lambda}_{(\rho\gamma\alpha)}+\epsilon_{\nu\gamma}^{\ \ \ \ \rho}\tilde{\Lambda}_{(\rho\alpha\beta)},
\end{eqnarray}
\no which indeed can be verified. We have also to say that the gauge parameter is symmetric and traceless.  However such transformation does not correspond to a symmetry of the action (\ref{s2}) because of the remaining terms. Then we can perform another round of the Noether procedure. which demands the calculation of the new Euler tensor:

\be M^{\mu(\beta\gamma\lambda)}=E^{\mu}_{\,\,\, \nu}\left[-2\Omega^{\nu(\beta\gamma\lambda)}(\xi)+m\omega^{\nu(\beta\gamma\lambda)}-\frac{2m}{3}\Omega^{\nu(\beta\gamma\lambda)}(f)\right]\equiv E^{\mu}_{\,\,\, \nu}\tilde{M}^{\nu(\beta\gamma\lambda)}. \label{M} \ee

As we have done before, we introduce an auxiliary field $b^{\nu(\beta\gamma\lambda)}$ coupled to the Euler tensor defining the first iterated action:

\begin{equation}
S'=S_{SD(2)}-\int d^3x \ \ [b^{\nu(\alpha\beta\gamma)}M_{\nu(\alpha\beta\gamma)}],
\end{equation}
where $b^{\nu(\alpha\beta\gamma)}$ has been choosen in order to have the same gauge transformation of the spin-4 field, $\delta_{\Lambda} \omega^{\nu(\beta\gamma\lambda)}=\delta_{\Lambda} b^{\nu(\beta\gamma\lambda)}$. Then, taking the gauge transformation of $S'$ we have:

\be \delta_{\Lambda} S'=\int d^3 x\ \ \delta_{\Lambda}\left[-\frac{m}{2} b^{\nu(\alpha\beta\gamma)}  E_\nu^{\ \ \lambda}b_{\lambda(\alpha\beta\gamma)}\right].\ee
Again, by construction, we have an invariant action given by:
\be S''= S_{SD(2)} -\int d^3x \ \  \left[b^{\nu(\alpha\beta\gamma)}E_\nu^{\ \ \lambda} \tilde{M}_{\lambda(\alpha\beta\gamma)}-\frac{m}{2}b^{\nu(\alpha\beta\gamma)}E_\nu^{\ \ \lambda} b_{\lambda(\alpha\beta\gamma)}\right].\ee

In order to eliminate the auxiliary field $b_{\lambda(\alpha\beta\gamma)}$ in terms of the Euler tensor we, differently of the previous case, notice that the last expression can be rewritten as:
\bea S''&=& S_{SD(2)} -\int d^3x \ \  \left[-\frac{m}{2}\left(b^{\nu(\alpha\beta\gamma)}- \frac{\tilde M^{\nu(\alpha\beta\gamma)}}{m}\right) E_\nu^{\ \ \lambda} \left(b_{\lambda(\alpha\beta\gamma)}- \frac{\tilde M_{\lambda(\alpha\beta\gamma)}}{m}\right)\right.\nn\\
&+&\left. \frac{1}{2m}\tilde{M}_{\lambda(\alpha\beta\gamma)}E_\nu^{\ \ \lambda} \tilde{M}^{\nu(\alpha\beta\gamma)}\right],\eea

\no which by shifting $b\to b +\tilde{M}/m$ completely decouple the auxiliary field from the Euler tensor. The remaining decoupled term on $b$ is a Chern-Simons like term which as we know has no particle content by itself and can be neglected from now on, leaving us with the following action:
\bea S''&=& S_{SD(2)} -\frac{1}{2m}\int d^3x \ \  \left[ \tilde{M}_{\lambda(\alpha\beta\gamma)}E_\nu^{\ \ \lambda} \tilde{M}^{\nu(\alpha\beta\gamma)}\right]. \eea

After the substitution of $\tilde{M}$ defined in (\ref{M}), making use of the properties of $\Omega$, and some integration by parts we have finally a third order in derivatives self-dual action for the spin-4 mode:

\bea S_{SD(3)}&=& \int d^3x \left[ \xi_{\rho(\alpha\beta\gamma)}\Omega^{\rho(\alpha\beta\gamma)}(\xi)- \frac{2}{m}\Omega_{\rho(\alpha\beta\gamma)}(\xi) E^{\rho}_{\,\,\,\nu}\Omega^{\nu(\alpha\beta\gamma)}(\xi) \right.\nn\\ &-& \left. \frac{1}{2} f_{\rho(\alpha\beta\gamma)}(\tilde{U}) E^{\rho}_{\,\,\,\nu}\Omega^{\nu(\alpha\beta\gamma)}(\xi)+\frac{21m}{80}\tilde{U}_{(\alpha\beta)}E^{\alpha}_{\,\,\,\gamma}\tilde{U}^{(\gamma\alpha)} + {\cal L}^2_{aux}  \right]\label{sd3}\eea

In this third order self-dual model, which is an original result of this work, we can observe that we still have the presence of the second-order term, but with a change of sign, which is typical of this kind of self-dual models connected by the dualization procedure. What is new now is the presence of a third order term, which is also typical, with structure given by $\sim \Omega E \Omega$; as we will see in the next section when we make the migration to a totally symmetric notation such term becomes a symmetrized curl of the totally symmetric field. It is also interesting to notice that the auxiliary lagrangian ${\cal L}_{aux}^2$  has gained a new contribution $\sim \tilde{U}E \tilde{U}$, and from now on we redefine it, i.e.: ${\cal L}_{aux}^2 \to {\cal L}_{aux}^3$. Last, we also mention that in order to become gauge invariant, the linking term has also changed.

\section{ ``Geometry'' from the dreinbein dependent models}

It is well known, see for example appendix A of \cite{Deseryang}, that the number of independent components of a totally symmetric field of rank-$s$ in $D$ dimensions is given by the combination:
\be C(D+s-1,s)=\frac{(D+s-1)!}{s!(D-1)!}\ee

\no where the tracelessness condition can be achieved by removing $C(D+s-3,s-2)$ from it, while double tracelessness condition by removing $C(D+s-5,s-4)$. Then one can easily verify that the number of independent components of $\omega_{\mu(\alpha\beta\gamma)}$ is given by $N= 3\left[C(5,3)-C(3,1)\right]=21$. Such $21$ components may be organized in terms of a totally symmetric double traceless field, through:

\be \omega _{\mu(\beta\gamma\lambda)}= \phi_{\mu\beta\gamma\lambda}+\frac{1}{3}\eta_{\mu(\beta}\phi_{\gamma\lambda)}-\frac{1}{3}\eta_{(\beta\gamma}\phi_{\lambda)\mu}+ c \,\, \epsilon_{\mu\rho(\beta} \tilde{\chi}^{\rho}_{\,\,\,\gamma\lambda)},\label{deco}\ee

\no where we have chosen the numerical coefficients in order to respect the tracelessness condition of the $\omega_{\mu(\beta\gamma\lambda)}$ field in the left hand side of the equation. Besides, such coefficients are adjusted in order to reproduce the second order term used for example in \cite{Deseryang}. Notice that the last coefficient $c$ is kept arbitrary, once our terms are invariant under (\ref{Sim2}). The parenthesis in this expression means unnormalized symetrization of the indices. On the right hand side of (\ref{deco}) one can check that the number of independent components are the sum of those given by the symmetric double traceless field $\phi_{\mu\beta\gamma\lambda}$, given by $C(6,4)-C(2,0)=14$, plus those of the symmetric traceless field $\tilde{\chi}_{\rho\gamma\lambda}$ given by $C(5,3)-C(3,1)=7$ \footnote{Maybe one could try another decomposition to the dreibein field in such a way that the totally symmetric field would be non double traceless, this would be an attempt of reproducing the structures found for example in \cite{ddalrs}.}.

We would like to construct geometrical objects, like the Einstein tensor, for the double traceless totally symmetric field. In order to do that one can mimic the nice procedure the authors have done in \cite{deserdamour} for the rank three field. Here, we aim to find an action invariant under the gauge transformation:
\be \delta_{\tilde{\xi}}\phi_{\mu\beta\gamma\lambda}= \p_{(\mu}\tilde{\xi}_{\beta\gamma\lambda)}\quad ; \tilde \xi_{\beta}=0.\ee

\no We look then for a ``Christoffel'' symbol transforming as a gradient of the gauge parameter as closely as possible. After that, we could aim to find gauge invariants. This is respected by the first order in derivative symbol: 
\be \Gamma_{\mu\nu\lambda\beta}^{(1)\alpha}\equiv \p_{(\mu}\phi_{\nu\lambda\beta)}^{\alpha}-\p^{\alpha}\phi_{\mu\nu\lambda\beta}\quad; \quad \delta_{\xi} \Gamma_{\mu\nu\lambda\beta}^{(1)\alpha}= 2 \,\,\p_{(\mu}\p_{\nu}\xi_{\lambda\beta)}^{\alpha},\ee

\no and as in the rank three case, one can also define a second order symbol, which, under arbitrary reparametrization transforms as a multigradient of the gauge parameter, i.e.:
\be \Gamma_{\mu\nu\lambda\beta}^{(2)\alpha\gamma}\equiv\p_{(\mu} \Gamma_{\nu\lambda\beta)\gamma}^{(1)\alpha}- 2\,\,\p_{\gamma} \Gamma_{\mu\nu\lambda\beta}^{(1)\alpha} \quad; \quad \delta_{\xi}\Gamma_{\mu\nu\lambda\beta}^{(2)\alpha\gamma}= 6\,\, \p_{(\mu}\p_{\nu}\p_{\lambda} \xi_{\beta)}^{\gamma\alpha}.\ee

\no Once the second order in derivative symbol is $\alpha\gamma$-traceless we can define the ``Ricci'' symbol from it:
\be \mathbb{R}_{\mu\nu\lambda\beta}\equiv \frac{1}{2}\Gamma_{\mu\nu\lambda\beta}^{(2)}= \Box \phi_{\mu\nu\lambda\beta}-\p_{(\mu}\p^{\alpha}\phi_{\alpha\nu\lambda\beta)}+\p_{(\mu}\p_{\nu}\phi_{\lambda\beta)}\quad; \quad \delta_{\tilde{\xi}}\mathbb{R}_{\mu\nu\lambda\beta}=0. \ee

\no As in the lower spin cases it is also convenient to define along with the ``Ricci'' tensor its trace, which is given by:
\be \mathbb{R}_{\lambda\beta}=2\left[\Box \phi_{\lambda\beta}-\p^{\mu}\p^{\alpha}\phi_{\mu\alpha\lambda\beta}+\frac{1}{2}\p_{(\beta}\p^{\alpha}\phi_{\alpha\lambda)}\right]\ee

\no With the Ricci and its trace we can finally define the second order in derivative Einstein tensor:
\be \mathbb{G}_{\mu\nu\lambda\beta}\equiv \mathbb{R}_{\mu\nu\lambda\beta}-\frac{1}{2}\eta_{(\mu\nu}\mathbb{R}_{\lambda\beta)}.\label{G}\ee

We have to remember that all such deductions and definitions are merely illustrative, once as we know, they are precisely those obtained by Fronsdal. Actually, the so called Einstein tensor here, is the rank four Fronsdal tensor for many authors. The fact is that once we have in hand all this definitions we can construct a second order term for the symmetric double traceless field, which is given by:

\bea \frac{1}{2} \int d^3x\,\,\, \phi_{\mu\nu\lambda\beta}\,\mathbb{G}^{\mu\nu\lambda\beta}(\phi)&=& \frac{1}{2}\int d^3x\,\,\, \left[\phi_{\mu\nu\lambda\beta} \Box \phi^{\mu\nu\lambda\beta} +4 (\p^{\mu} \phi_{\mu\nu\lambda\beta})^2+12\phi_{\mu\nu} \p_{\lambda}\p_{\beta} \phi^{\mu\nu\lambda\beta}\right.\nn\\ &-&\left.6\phi_{\mu\nu} \Box \phi^{\mu\nu} -6 \phi_{\mu\nu} \p^{\mu}\p_{\alpha} \phi^{\nu\alpha}\right] .\eea

\no Which is precisely the second order term (see expression 4.2)  of \cite{Deseryang}. It is also useful to notice that the operator $\mathbb{G}^{\mu\nu\lambda\beta}(\phi)$ is self-adjoint in the sense that $\int \psi \mathbb{G}(\phi)=\int \phi \mathbb{G}(\psi)$ . Such property will be determinant for obtaining equations of motion as well as for the interpolation with other self-dual models.

Once we are dealing with parity violating actions, it is useful to define from expression 5.1 of \cite{deserdamour} a symmetrized curl: 
\be \mathbb{C}_{\mu\nu\gamma\lambda}(\phi)\equiv - E_{(\mu}^{\,\,\,\,\,\,\beta}\phi_{\beta\nu\gamma\lambda)},\label{C}\ee
which by its turn allow us to construct a third order gauge invariant term given by:

\bea \int d^3x\,\,\mathbb{C}_{\mu\nu\gamma\lambda}(\phi)\mathbb{G}^{\mu\nu\gamma\lambda}(\phi)&=& \int d^3x\,\, \left[-4 \phi_{\beta\nu\gamma\lambda}\Box E_{\s\mu}^{\beta} \phi^{\mu\nu\gamma\lambda} + 12 \phi_{\beta\mu\gamma\lambda}E_{\s\nu}^{\beta}\p^{\mu}\p_{\alpha}\phi^{\alpha\nu\gamma\lambda}\right.\nn\\
&-&\left.24\phi_{\beta\mu\nu\lambda}E_{\s\gamma}^{\beta}\p^{\mu}\p^{\nu}\phi^{\gamma\lambda}+12\phi_{\beta\lambda}\Box E_{\s\gamma}^{\beta} \phi^{\gamma\lambda}+6\phi_{\beta\lambda}E_{\s\gamma}^{\beta}\p^{\lambda}\p_{\alpha}\phi^{\alpha\gamma}\right].\label{C4}\nn\\\eea

Similar to before, the operator $\mathbb{C}_{\mu\nu\gamma\lambda}(\phi)$ is also self-adjoint and we are going to need that property in the next section, i.e.:  $\int \psi \mathbb{C}(\phi)=\int \phi \mathbb{C}(\psi)$.  Besides, one can also check that the operators $\mathbb{C}$ and $\mathbb{G}$ commute each other $\int \phi \mathbb{C}[\mathbb{G}(\phi)]=\int \phi \mathbb{G}[\mathbb{C}(\phi)]$.

As a final comment of this section we are going to verify that our $\omega$-dependent terms obtained at (\ref{sd3}) can be translated to the symmetric notation via the decomposition (\ref{deco}). Below, we give the explicit expressions for our terms as well as their geometrical forms:
\bea \int d^3x\,\, \xi_{\rho(\alpha\beta\gamma)}\Omega^{\rho(\alpha\beta\gamma)}(\xi)&=&-\frac{7}{8}\omega_{\mu(\beta\gamma\lambda)}\Box \omega^{\mu(\beta\gamma\lambda)}+\frac{7}{8}\omega_{\mu(\beta\gamma\lambda)}\p^{\mu}\p_{\nu}\omega^{\nu(\beta\gamma\lambda)}+\frac{3}{2}\omega_{\mu\beta}\Box \omega^{\mu\beta}\nn\\
&+&\frac{15}{8}\omega_{\mu(\beta\gamma\lambda)}\p^{\beta}\p_{\nu}\omega_{\mu(\nu\gamma\lambda)}-3\omega_{\mu(\beta\gamma\lambda)}\p^{\mu}\p^{\beta}\omega^{\gamma\lambda}+\frac{3}{8}\omega_{\mu(\beta\gamma\lambda)}\Box \omega^{\beta(\mu\gamma\lambda)}\nn\\
&-&\frac{3}{4}\omega_{\mu(\beta\gamma\lambda)}\p^{\mu}\p_{\nu}\omega^{\beta(\nu\gamma\lambda)}\nn\\
&=& \frac{1}{2} \int d^3x\,\,\, \phi_{\mu\nu\lambda\beta}\,\mathbb{G}^{\mu\nu\lambda\beta}(\phi),\eea
\no for the second order term,
\bea -\frac{2}{m}\int d^3x\,\, \Omega_{\rho(\alpha\beta\gamma)}(\xi) E^{\rho}_{\,\,\,\nu}\Omega^{\nu(\alpha\beta\gamma)}(\xi)&=&\omega_{\lambda(\alpha\beta\gamma)}\Box E^{\lambda}_{\s\rho}\omega^{\rho(\alpha\beta\gamma)}-\frac{3}{2}\omega_{\lambda(\alpha\beta\gamma)}\Box E^{\alpha}_{\s\rho}\omega^{\rho(\lambda\beta\gamma)}\nn\\
&+&\frac{3}{2}\omega_{\lambda(\alpha\beta\gamma)}E^{\alpha}_{\s\rho}\p^{\lambda}\p_{\nu}\omega^{\rho(\nu\beta\gamma)}-\frac{3}{32}\omega_{\beta\gamma}\Box E_{\rho\sigma}\omega^{\sigma(\rho\beta\gamma)}\nn\\
&+&\frac{3}{32}\omega_{\lambda(\alpha\beta\gamma)}\p^{\lambda}\p^{\alpha} E_{\rho\sigma}\omega^{\sigma(\rho\beta\gamma)}+\frac{3}{4}\omega_{\lambda(\mu\beta\gamma)}\Box E^{\mu}_{\s\nu}\omega^{\lambda(\nu\beta\gamma)}\nn\\
&-&\frac{3}{4}\omega_{\lambda(\mu\beta\gamma)} E^{\mu}_{\s\nu}\p^{\lambda}\p_{\rho}\omega^{\rho(\nu\beta\gamma)}-\frac{3}{2}\omega_{\lambda(\mu\beta\gamma)}\Box E^{\beta}_{\s\rho}\p^{\mu}\p_{\alpha}\omega^{\rho(\lambda\alpha\gamma)}\nn\\
&+&\frac{3}{2}\omega_{\beta\gamma}\Box E^{\beta}_{\s\rho}\p_{\alpha}\p_{\nu}\omega^{\rho(\nu\alpha\gamma)}
+\frac{33}{32}\omega_{\gamma(\lambda\mu\beta)}\p^{\lambda}\p^{\mu}E_{\rho\sigma}\omega^{\sigma(\rho\gamma\beta)}\nn\\
&-&\frac{33}{32}\omega_{\lambda\beta}\p^{\lambda}\p_{\nu}E_{\rho\sigma}\omega^{\sigma(\rho\nu\beta)}\nn\\ 
&=& \frac{1}{8m} \int d^3x\,\,\mathbb{C}_{\mu\nu\gamma\lambda}(\phi)\mathbb{G}^{\mu\nu\gamma\lambda}(\phi)\eea
\no for the third order term, and finally, the linking term, given by:
\bea -\frac{1}{2}\int d^3x\,\, f_{\rho(\alpha\beta\gamma)}(\tilde{U}) E^{\rho}_{\,\,\,\nu}\Omega^{\nu(\alpha\beta\gamma)}(\xi)&=& -\frac{9}{8}\int d^3x\,\,\tilde{U}_{(\beta\gamma)}\Box \omega^{\beta\gamma}+\frac{9}{8}\tilde{U}_{\beta\gamma}\p_{\alpha}\p_{\lambda}\omega^{\lambda(\alpha\beta\gamma)}\nn\\
&+&\frac{9}{8}\tilde{U}_{(\beta\gamma)}\p_{\nu}\p_{\alpha}\omega^{\beta(\nu\alpha\gamma)}-\frac{9}{8}\tilde{U}_{(\beta\gamma)}\p^{\beta}\p_{\alpha}\omega^{\alpha\gamma}\nn\\ &=&\frac{3}{40}\eta_{(\mu\nu}\tilde{U}_{\lambda\beta)}\mathbb{G}^{\mu\nu\lambda\beta}(\phi).\eea

Once all the $\omega$-dependent terms can be translated, to the geometrical description, which means that they can be written in terms of $\mathbb{G}$ one can wonder about the possibility of finding a new self-dual model of fourth order in derivative as it was the case in the spin-3 context. Finally, we have to point out a difference of notation here. Under the approach we have used here, based on \cite{deserdamour}, the Einstein tensor is second order in derivatives, however in \cite{henneaux} the authors have done a systematically study on the conformal geometry of higher spin bosonic gauge fields in three spacetime dimensions where the Einstein tensor is proportional to the Riemann tensor which for a rank-$s$ field is of order $s$ in derivatives. 

\section{From $SD(3)$ to $SD(4)$}
In order to get a new self-dual model of fourth order in derivatives from the third order one, we have to investigate the symmetries encoded in the third order term which are absent in the second and even in the linking term. One could use the spin-3 case as an example: When passing from the $SD(3)$ to $SD(4)$ in that case, see \cite{nges3} we suggested a generalization of the traceless diffeomorphism  where the gauge parameter became trace full. The analogue of this, here, would be $\delta_{\xi} \phi_{\mu\nu\lambda\beta}=\p_{(\mu}\xi_{\nu\lambda\beta)}$, but the circumstances are a bit more subtle than this. Once the field must be double traceless according to our prescription given by (\ref{deco}) we have to suggest a transformation slightly different and unusual:
\be \delta_{\xi} \phi_{\mu\nu\lambda\beta}=\p_{(\mu}\xi_{\nu\lambda\beta)}-\frac{4}{15}\eta_{(\mu\nu}\eta_{\lambda\beta)}\p^{\alpha}\xi_{\alpha}. \label{sym4}\ee

\no Using the explicit expression for the third order term given by (\ref{C4}), we can check that such transformation (\ref{sym4}) will become a symmetry of this term, if and only if we have the trace of the gauge parameter as a longitudinal vector, in other words:
\be \xi_{\mu}=\p_{\mu}\psi\quad; \quad \psi=\psi(x). \label{cond4}\ee

Using the explicit expression for the second order term, it also easy to check that (\ref{sym4}) with the additional condition (\ref{cond4}) does not configure a symmetry of that term, which then allow us to try another round of Noether gauge embedment. Perhaps would be interesting to notice that the double Weyl part of (\ref{sym4}), the term  given by $\sim \eta \eta \p\cdot \xi$ is by itself a symmetry even of the third as the second order term.
We start by writing the $SD(3)$ model in its geometrical form:
\be S_{SD(3)}= \int d^3x \left[\frac{1}{2} \phi_{\mu\nu\lambda\beta}\,\mathbb{G}^{\mu\nu\lambda\beta}(\phi) +\frac{1}{8m}\mathbb{C}_{\mu\nu\gamma\lambda}(\phi)\mathbb{G}^{\mu\nu\gamma\lambda}(\phi)+{\cal U}_{\mu\nu\lambda\beta}\mathbb{G}^{\mu\nu\lambda\beta}(\phi)+{\cal L}_{aux}^3\right]\ee
\no where we have defined ${\cal U}_{\mu\nu\lambda\beta}\equiv 3\eta_{(\mu\nu}\tilde{U}_{(\lambda\beta))} /40$. Then, taking advantage that all the terms in the action are proportional to the operator $\mathbb{G}$, it can be factored and the equations of motion with respect to the totally symmetric field $\phi_{\mu\nu\lambda\beta}$ are quite simple:

\be N^{\mu\nu\lambda\beta}= \mathbb{G}^{\mu\nu\lambda\beta}\left[ \phi + \frac{1}{4m}\mathbb{C}(\phi)+{\cal U}\right]\equiv N^{\mu\nu\lambda\beta} (b).\label{B}\ee

\no Where we have defined $b \equiv  \phi + \frac{1}{4m}\mathbb{C}(\phi)+{\cal U}$.  Next, we proceed by suggesting an auxiliary field $a_{\mu\nu\lambda\beta}$ transforming in the same way the original field does, i.e.: $\delta_{\xi} a_{\mu\nu\lambda\beta}=\delta_{\xi}\phi_{\mu\nu\lambda\beta}$ which allow us to conclude that:
\be S_{1}= S_{SD(3)}-\int d^3x\,\, a_{\mu\nu\lambda\beta} N^{\mu\nu\lambda\beta}\quad,\quad \delta_{\xi}S_1= -\frac{1}{2} \int d^3x\,\, \delta(a_{\mu\nu\lambda\beta}\mathbb{G}^{\mu\nu\lambda\beta}(a)).\ee

By construction, we get then a $\xi$-gauge invariant action given by:

\be S_2= S_{SD(3)}-\int d^3x \,\, \left[ a_{\mu\nu\lambda\beta}\mathbb{G}^{\mu\nu\lambda\beta}(b)-\frac{1}{2}a_{\mu\nu\lambda\beta}\mathbb{G}^{\mu\nu\lambda\beta}(a)\right]. \label{above}\ee

The auxiliary field can be easily eliminated, if we notice that the term under the integral (\ref{above}) can be written as:
\be S_2= S_{SD(3)}+\frac{1}{2}\int d^3x \,\, \left[ (a-b)_{\mu\nu\lambda\beta}\mathbb{G}^{\mu\nu\lambda\beta}(a-b)\right] -\frac{1}{2}\int d^3x \,\,\left[b_{\mu\nu\lambda\beta}\mathbb{G}^{\mu\nu\lambda\beta}(b)\right]. \label{free}\ee

\no Considering that the second term in (\ref{free}) is free of particle content and completely decoupled from the rest of the action by performing $a \to a +b$, we end up with:
\be S_2= S_{SD(3)} -\frac{1}{2}\int d^3x \,\,\left[b_{\mu\nu\lambda\beta}\mathbb{G}^{\mu\nu\lambda\beta}(b)\right]. \label{free2}\ee

Substituting back $b$, defined in (\ref{B}), we finally get the fourth order, gauge invariant action:
\bea S_{SD(4)}&=& \int d^3x \,\,\left[-\frac{1}{8m}\mathbb{C}_{\mu\nu\gamma\lambda}(\phi)\mathbb{G}^{\mu\nu\gamma\lambda}(\phi)-\frac{1}{32m^2}\mathbb{C}_{\mu\nu\gamma\lambda}(\phi)\mathbb{G}^{\mu\nu\gamma\lambda}(\mathbb{C})-\frac{1}{4m}\mathbb{C}_{\mu\nu\gamma\lambda}(\phi)\mathbb{G}^{\mu\nu\gamma\lambda}({\cal U})\right.\nn\\
&-& \left. \frac{1}{2}{\cal {U}}_{\mu\nu\gamma\lambda}\mathbb{G}^{\mu\nu\gamma\lambda}({\cal{U}})+  {\cal L}_{aux}^3\right].\eea

The self-dual massive fourth order model we have obtained here is complete, in the sense that, it contains all the auxiliary fields needed to describe the unique spin-4 mode. As an automatic consequence of the procedure we have used, the auxiliary action has been corrected once more, with:
\be - \frac{1}{2}{\cal {U}}_{\mu\nu\gamma\lambda}\mathbb{G}^{\mu\nu\gamma\lambda}({\cal{U}})=\frac{81}{320}\left(2 \tilde{U}_{(\mu\nu)}\Box \tilde{U}^{(\mu\nu)}-\tilde{U}_{(\mu\nu)}\p^{\mu}\p_{\alpha}\tilde{U}^{(\alpha\nu)}\right).\ee
 
\no By incorporating this correction to the previous auxiliary action we have ${\cal L}_{aux}^3 \to {\cal L}_{aux}^4$. Besides, the new gauge invariant third order linking  term $\sim {\cal U}\mathbb{G}(\mathbb{C})$ is explicitly given by:
\be -\frac{1}{4m}\mathbb{C}_{\mu\nu\gamma\lambda}(\phi)\mathbb{G}^{\mu\nu\gamma\lambda}({\cal U})=-\frac{9}{16m}\left[ 2\phi_{\mu\nu}\Box E^{\mu}_{\s\alpha}\tilde{U}^{(\alpha\nu)}-2\phi_{\mu\nu\gamma\lambda}E^{\mu}_{\s\beta}\p^{\nu}\p^{\gamma}\tilde{U}^{(\beta\lambda)}+\phi_{\mu\nu}E^{\mu}_{\s\alpha}\p^{\nu}\p_{\beta}\tilde{U}^{(\alpha\beta)}\right].\ee
Finally, we have generated the fourth order term:
\bea -\frac{1}{32m^2}\mathbb{C}_{\mu\nu\gamma\lambda}(\phi)\mathbb{G}^{\mu\nu\gamma\lambda}(\mathbb{C})&=& -\frac{1}{2}\phi_{\mu\nu\gamma\lambda}\Box^2\phi^{\mu\nu\gamma\lambda}+2 \phi_{\mu\nu\gamma\lambda}\Box \p^{\mu}\p_{\alpha}\phi^{\alpha\nu\gamma\lambda}\nn\\
&-&\frac{9}{4}\phi_{\mu\nu\gamma\lambda}\Box\p_{\mu}\p^{\nu}\phi^{\gamma\lambda}-\frac{15}{8}\phi_{\mu\nu\gamma\lambda}\p^{\mu}\p^{\nu}\p_{\alpha}\p_{\beta}\phi^{\alpha\beta\gamma\lambda}\nn\\
&+&\frac{9}{8}\phi_{\mu\nu}\Box^2\phi^{\mu\nu}-\frac{15}{16}\phi_{\mu\nu}\p^{\mu}\p^{\nu}\p_{\alpha}\p_{\beta}\phi^{\alpha\beta}\nn\\
&+&\frac{15}{4}\phi_{\mu\nu\gamma\lambda}\p^{\mu}\p^{\nu}\p_{\gamma}\p_{\alpha}\phi^{\alpha\lambda}-\frac{27}{16}\phi_{\mu\nu}\Box\p^{\mu}\p_{\alpha}\phi^{\alpha\nu},\label{4t}\eea

\no which is invariant under all the previous gauge transformations. It is also interesting to notice that all the results we have obtained here are quite similar to those we have found in the spin-3 case. As in that case, we have also observed that, as far as we can investigate there is no gauge transformation to implement into the fourth order model. In other words, the fourth order term does not have an invariance under a new symmetry which would be broken by the third order term. One can also notice that the fourth order term we have obtained here does not correspond to the fourth order term suggested in the very interesting thesis \cite{marijat, marija} where the author has defined a generalized fourth order Einstein tensor, in such a way that ${\cal L}^{(4)} = \phi_{\mu\nu\lambda\beta} E_{\mu}^{\s\alpha}E_{\nu}^{\s\gamma}E_{\lambda}^{\s\sigma}E_{\beta}^{\s\rho}\phi_{\alpha\gamma\sigma\rho}$. In fact, that term is invariant under unconstrained gauge transformation $\delta \phi_{\mu\nu\lambda\beta}= \p_{(\mu}\xi_{\nu\lambda\beta)}$ which is broken by (\ref{4t}). It has been quite challenging to overcome this barrier, in order access the highest order models, for example those obtained in \cite{ddalrs}, perhaps this will remain impossible once even those highest order models are unconnected ab initio.

\section{Conclusion}
In this work we have used a spin-4 self-dual model in $D=2+1$ dimensions suggested for the first time in \cite{aragones31}, as a starting point to obtain a sequence of three more self-dual  descriptions which are of second, third and fourth order in derivatives. All of them are achieved through a dualization procedure we have similarly used in lower spin cases for both, bosons and fermions, the procedure is called the Noether Gauge Embedment.

It turns out that the descriptions of first, second and third order in derivatives are written in terms of dreibein fields $\omega_{\mu(\alpha\beta\gamma)}$ and the complicated terms of higher derivatives as well as the procedure by itself, becomes manageable thanks to a powerful notation which introduces the self-adjoint operator $\Omega_{\mu(\alpha\beta\gamma)}$ (\ref{OM}). Similar structure has already been used in the lower spin bosonic cases, spin-2 and 3. Once the spin-4 field is the very first representative of genuine higher spin descriptions (due to the double traceless condition) we believe that further generalizations of such operator for higher spins, i.e.: $\Omega_{\mu_1(\mu_2 ... \mu_s)}$ can be suggested having this case as an example. 

We have demonstrated that, once we get to the third order self-dual description, it is possible a conversion of the frame-like notation in  terms of a ``geometrical" one, which is entirely described in terms of a totally symmetric field $\phi_{\mu\nu\lambda\beta}$. Such geometrical description is reached out by extending the steps of \cite{deserdamour} for the case of rank four fields, which allow us to obtain the so called Einstein tensor $\mathbb{G}_{\mu\nu\lambda\beta}$ (\ref{G})  and symmetrized-curl $\mathbb{C}_{\mu\nu\lambda\beta}$ (\ref{C}), both symmetric. Such operators enjoy of useful algebraically properties such as self-adjointness and commutation and this is precisely why we have been able to obtain a fourth order self-dual model. The last self-dual model is invariant under the complete set of gauge transformations implemented along the procedure, i.e. (\ref{sim1}), (\ref{Sim2}) and (\ref{sym4}). As far as we can investigate, there is no new gauge transformation to be imposed, and again, as we have verified in the much simpler case of spin-3, we are stuck in the fourth stage.   

Some directions must be better investigated after this work. We have observed that, once we have the auxiliary fields established in the very first self-dual model $SD(1)$, the subsequent models receive step by step new corrections and new gauge invariant linking terms substituting the non gauge invariant ones. A next and important step would be the consideration of a source term coupled to the spin-4 field in $SD(1)$. From this source-field coupling we can obtain the corresponding dual maps among the descriptions, connecting for example the equations of motion in the classical level. Besides and much more interesting and important, with such dual maps, one can also verify the quantum equivalence between the suggested models via a unique master action interpolating among the descriptions by comparing correlation function calculated for $N$ points.  In this sense the $\Omega$ notation is a technical prerequisite in order to construct such master action, besides, one has to demonstrate the absence of particle content of the so called mixing terms and this seems to be tricky for example for the third order term. Finally, we hope to be able to develop the $\Omega$ notation in order to find out higher derivative versions from the arbitrary massive spin-s bosonic and fermionic actions introduced by \cite{Tyutin}. Another interesting point would be the study of the dual map connecting the models given by the expressions (4.5b and 7.1) by \cite{Kuzenko} which are higher spin analogues of the self-dual descriptions we have studied through the systematic Noether Gauge Embedment approach.

\section{Acknowledgements}

The authors would like to thank Prof. Denis Dalmazi for useful discussion regarding geometry and gauge symmetries. The work of H.L.O has been supported by CAPES.

\end{document}